\documentclass[pdflatex,sn-mathphys-num,iicol]{sn-jnl}

\usepackage[numbers]{natbib}
\usepackage{amsmath}

\usepackage[mathscr]{euscript}
\usepackage{amssymb}

\newcommand{\GeV}{{\,\mathrm{GeV}}}
\newcommand{\rd}{{\mathrm{d}}}

\newcommand{\imag}{\mathfrak{Im}}
\newcommand{\sgn}{{\rm sgn}}

\newcommand{\p}{{\bf p}}
\newcommand{\BE}{f_{\rm B}}
\newcommand{\FD}{f_{\rm F}}
\newcommand{\fBF}{f_{\rm B/F}}
\newcommand{\ie}{i.e.}
\newcommand{\pv}{\mathcal{P}}
\newcommand{\sftype}[1]{{\mathrm{t}(#1)}}
\newcommand{\red}[1]{{\color{black}#1}}

\begin{document}
\allowdisplaybreaks

\title[CP asymmetry factor in decays at finite temperature]{CP asymmetry factor in decays at finite temperature}

% authors

\author*[1,2]{\fnm{Károly} \sur{Seller}}
\email{karoly.seller@uni-hamburg.de}

\author[3]{\fnm{Zsolt} \sur{Szép}}
\email{szepzs@achilles.elte.hu}

\author[1,3]{\fnm{Zoltán} \sur{Trócsányi}}
\email{zoltan.trocsanyi@cern.ch}


\affil*[1]{
\orgdiv{Department of Theoretical Physics},
\orgname{ELTE Eötvös Loránd University},
\orgaddress{Pázmány Péter sétány 1/A},
\city{Budapest},
\postcode{1117},
\country{Hungary}}

\affil[2]{
\orgdiv{II. Institut für Theoretische Physik},
\orgname{Universität Hamburg},
\orgaddress{Luruper Chaussee 149},
\city{Hamburg},
\postcode{22761},
\country{Germany}}

\affil[3]{
\orgname{HUN-REN ELTE Theoretical Physics Research Group},
\orgaddress{Pázmány Péter sétány 1/A},
\city{Budapest},
\postcode{1117},
\country{Hungary}}

% abstract

\abstract{
We present in the conventional equilibrium approach to leptogenesis the complete leading order prediction for the CP asymmetry factor in finite-temperature decays involving Majorana neutrinos.
As thermal effects are generally not negligible, the knowledge of the high temperature behavior of the underlying particle physics model, in particular that of mass generation, is required for reliable estimates of matter-antimatter asymmetry through the mechanism of leptogenesis.
We present all necessary information needed to obtain the full one-loop evaluation for the thermal CP asymmetry factors for the processes $N_i\to \phi + L$ and $\phi \to N_i+L$ at temperatures where they are kinematically allowed.
We present a numerical comparison with previous formulae given in the literature.
}


\maketitle

\section{Introduction}

Decays involving heavy Majorana fermions could play an important role in the generation of the baryon asymmetry observed in the present Universe through the mechanism of leptogenesis \cite{Fukugita:1986hr}.
The measure of lepton number violation is usually expressed with the CP asymmetry factor, which is proportional to the difference between the rates of decays producing leptons and antileptons.
For a complete description of CP violation in leptogenesis via the Boltzmann equations, the CP asymmetry factor from decays involving right-handed neutrinos is supplemented by scattering processes \red{\cite{Biondini:2017rpb,Laine:2022pgk}}, which attempt to maintain equilibrium and wash out any generated asymmetry, as well as flavor effects \red{(the latter become important at lower temperatures). 
The actual matter-antimatter asymmetry and effective CP-violation factor also depend on the initial state of leptogenesis \cite{Klaric:2021cpi}. 
For instance, it is strongly influenced by whether the heavy right-handed neutrinos are in thermal equilibrium, or by the presence of oscillations that significantly mix neutrino flavors.}
In this letter we focus only on the CP asymmetry factor in the unflavored case.

We consider a Yukawa-type interaction between Majorana  %(right-handed) 
neutrinos, leptons, and a scalar field.
Assuming that the Yukawa coupling $y$ associated with this vertex is complex, %valued, 
the expression at leading order and at zero temperature for vanishing scalar and lepton masses is
\footnote{We use natural units defined by $\hbar=k_\mathrm{B}=c=1$.} 
\cite{Liu:1993tg,Xing:2011zza}
\begin{eqnarray}
    \label{eq:zeroTepsilon}
    &&\epsilon^{(0)}_{N_i\to \phi+L}(x)= \\
    &&\quad\frac{G}{8\pi}\sqrt{x}\Big\{\frac{1}{1-x} + \Big[1 + (1+x)\log\Big(\frac{x}{1+x}\Big)\Big]\Big\}\,,
\nonumber
\end{eqnarray}
where we used $G=\imag[(\mathcal{K}_{ij})^2]/\mathcal{K}_{ii}\neq 0$ with $\mathcal{K}=y^\dagger y$, and $x=m_{N_j}^2/m_{N_i}^2>1$ is the ratio of the squared masses of the Majorana neutrinos.
Eq.~\eqref{eq:zeroTepsilon} is obtained 
%, using natural units defined by $\hbar=k_\mathrm{B}=c=1$, 
as the sum of two one-loop contributions to the neutrino decay: (i) the one-loop neutrino self-energy diagram (the first term in curly brackets) and (ii) the vertex (triangle) diagram (in square brackets).
For strongly hierarchical neutrino masses ($x\gg 1$) the self-energy contribution is approximately twice that of the vertex contribution.

In the early Universe, the rate of CP violation should be calculated taking into account thermal effects due to the cosmic plasma, as first pointed out in Ref.~\cite{Covi:1997dr} and later expanded on by the influential article on thermal leptogenesis of Ref.~\cite{Giudice:2003jh}.
In this letter, following Ref.~\cite{Seller:2024dkg}, we present the full thermal contribution to the CP asymmetry factor (also called measure of CP violation below) in both $N_i\to \phi +L$ and $\phi\to N_i+L$ decays.
These are the functions to be used in phenomenological studies of thermal leptogenesis in the Boltzmann equation. 
The content of our formulae is not new in the sense that it can be collected from the non-equilibrium literature by cumbersome research and analytic work (see  discussion below Eq.~\eqref{eq:strippedMvertex3}).
However, the formulation presented here using explicit information from Ref.~\cite{Seller:2024dkg} is new and we believe that summarizing it in the way we present below will be valuable in the astroparticle literature, where  it is common to use a vacuum approximation for the CP asymmetry factor that neglects the scalar and lepton masses even today.

There exist two distinct methods for evaluating the lepton asymmetry generated in the early Universe: a more phenomenological bottom-up approach based on the semi-classical Boltzmann equation and a top-down functional method based on the Kadanoff-Baym equations \cite{Beneke:2010wd,Anisimov:2010aq}. 
\red{The latter is a consistent quantum description of non-equilibrium processes.
For a review of the progress made with this approach, see ref.~\cite{Garbrecht:2018mrp}.
A comparison of Kadanoff–Baym and Boltzmann results shows that the latter approach overestimates the amount of resonant enhancement of the lepton asymmetry in the limit of extreme mass degeneration \cite{Garny:2011hg}.}

\red{Initially, the non-equilibrium functional method led to a different measure of CP violation than the semi-classical (equilibrium) approach within the common range of applicability \cite{Garny:2009qn,Garny:2009rv}.}
This contradiction was resolved in Ref.~\cite{Garny:2010nj} where it was shown that the predictions for the CP asymmetries coincide in a scalar toy model provided that in the equilibrium approach one considers the causal $n$-point function (originally introduced by Kobes \cite{Kobes:1990kr}).
The discrepancy was due to an incorrect identification of the physical quantity in the bottom-up calculation and \emph{not} to non-equilibrium effects. 
Later works concerned with non-equilibrium thermal field theory expanded on Ref.~\cite{Garny:2010nj} by involving all sources of CP violation, in particular by considering all cuts of the two-loop Majorana neutrino two-point function (particularly those cuts of the vertex diagram that involve an on-shell neutrino) \cite{Garbrecht:2010sz,Frossard:2012pc}.

From the equilibrium field theory side, the measure of CP violation in decays can be obtained using either the imaginary-time or the real-time formalisms \cite{Kobes:1990kr,Aurenche:1991hi}.
In Ref.~\cite{Seller:2024dkg} we presented a more accessible semi-analytic method (discussed here in some detail in Section~\ref{section:2}) for the evaluation of the integrals appearing in the $N_i\to\phi+L$ decay, in particular for the purely thermal cuts of the vertex function.
In this letter we used the same method to calculate both the self-energy and the purely thermal vertex cut contributions also to $\epsilon_{\phi\to N_i+L}$. 
In principle, $\epsilon_{N_i\to\Phi+L}$ and $\epsilon_{\phi\to N_i+L}$ can be obtained also using Kadanoff-Baym approach, as discussed below Eq.~\eqref{eq:strippedMvertex3}.

The thermal rate for a process $a\to b+c$ is defined as ($y_p=E_p/T$):
\begin{eqnarray}
\label{eq:thermal_rate_definition}
%\begin{aligned}
\gamma_{a\to b+c} &&= \int\rd\Pi_{a}\rd\Phi_{\{b,c\}}\langle|\mathcal{M}_{a\to b+c}|^2\rangle \\
&&\times f_{\sftype{a}}(y_a) \big[1\pm f_{\sftype{b}}(y_b)\big]\big[1\pm f_{\sftype{c}}(y_c)\big]\,,
\nonumber
%\end{aligned}
\end{eqnarray}
where $\rd\Phi_{\{b,c\}}=\rd\Pi_b\rd\Pi_c (2\pi)^4\delta^{(4)}(P_a-P_b-P_c)$ is the two-particle Lorentz-invariant phase space measure, $\mathcal{M}_{a\to b+c}$ is the thermal amplitude associated to the process ${a\to b+c}$, and $\sftype{p}=$ B(ose) or F(ermi) gives the statistics type of particle $p$.
Clearly, $\gamma_{a\to b+c}$ violates Lorentz-invariance explicitly because of the statistical factors $\fBF(y)=[\exp(y)\mp1]^{-1}$.

Denoting the CP conjugate state of a particle $p$ as $\bar p$, the CP asymmetry factor is:
\begin{equation}
    \label{eq:epsilonT_definition}
    \epsilon_{a\to b+c} = \frac{\gamma_{a\to b+c} - \gamma_{\Bar{a}\to \Bar{b} + \Bar{c}}}{\gamma_{a\to b+c} + \gamma_{\Bar{a}\to \Bar{b} + \Bar{c}}}\,.
\end{equation}
In the processes we consider, one external particle is a Majorana fermion for which $\bar p = p$ and thus, the other two states may only be a fermion and a boson (no other 3-point vertex is conceivable).

\begin{figure*}[!ht]
\centering
\includegraphics[width=0.6\linewidth]{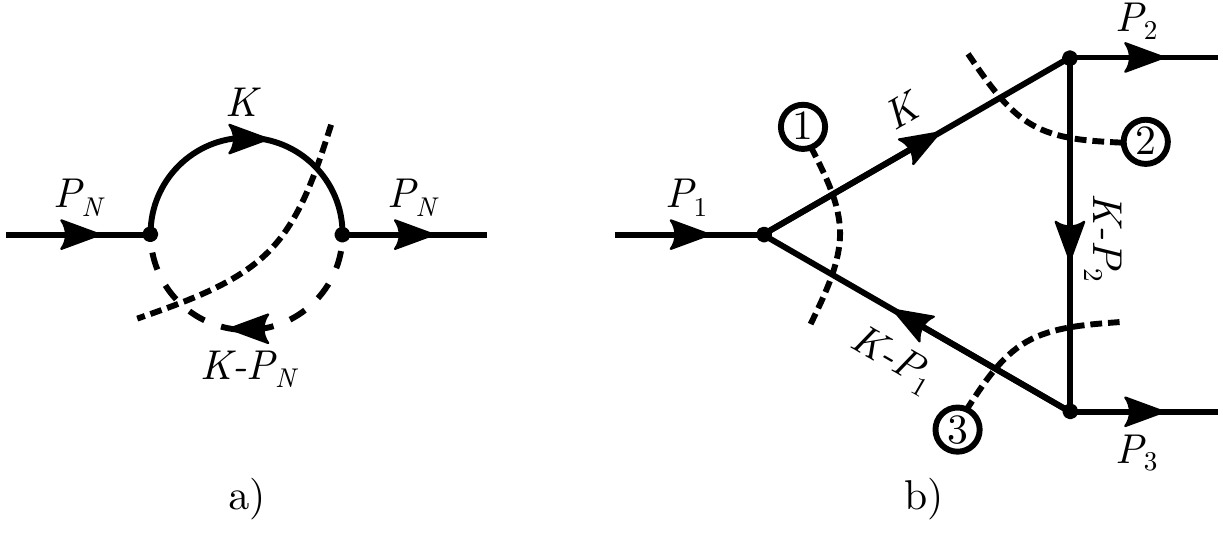}
\caption{One-loop cut diagrams contributing to the CP asymmetry factor.
a) The cut neutrino self-energy (bubble) diagram with intermediate lepton and scalar fields in the loop. 
b) The three cuts of the one-loop vertex correction (triangle) diagram. 
The particle types of the incoming and outgoing states as well as the internal lines depend on the particular process considered, $N_i(P_1)\to \phi(P_2) + L(P_3)$ or $\phi(P_1)\to N_i(P_2) + L(P_3)$.}
\label{fig:diagrams_combined}
%\vspace*{-20pt}
\end{figure*}

The CP asymmetry factor given in Eq.~\eqref{eq:epsilonT_definition} is evaluated here at leading order in perturbation theory.
At tree level the amplitude is equal to its CP conjugate, thus at the leading non-vanishing order the numerator is given by the interference between the tree level and one loop diagrams, whereas the denominator is simply twice the tree level decay rate.
With $x = \cos\theta$ and $z_{a} = m_{a}/T$, the CP asymmetry factor is given as the thermal average
\newpage
\begin{strip}
\begin{equation}
\label{eq:thermal_average}
\epsilon_{a\to b+c} = \frac{\displaystyle \int_{z_{a}}^\infty\!\rd y_a \, f_{\sftype{a}}(-y_a)\sqrt{y_a^2-z_{a}^2} \int_{-1}^1 \! \rd x \, \epsilon_{\mathcal{M}}(y_a,x) f_{\sftype{b}}(y_b)f_{\sftype{c}}(y_c)}{\displaystyle \int_{z_{a}}^\infty \! \rd y_a \, f_{\sftype{a}}(-y_a) \sqrt{y_a^2-z_{a}^2} \int_{-1}^1 \! \rd x \, f_{\sftype{b}}(y_b)f_{\sftype{c}}(y_c)}\,,
\end{equation}
\end{strip}
\noindent
where $\epsilon_{\mathcal{M}}$ is the amplitude-level CP asymmetry factor (see, Ref.~\cite{Giudice:2003jh,Seller:2024dkg}).
It is proportional to the imaginary part of the amplitude, while its form depends on the specific process and particle physics model.
For processes involving the Yukawa interaction between the Majorana neutrinos, leptons, and scalars, the CP asymmetry at leading order is \cite{Giudice:2003jh,Seller:2024dkg}:
\begin{equation}
\label{eq:epsilon_M}
\epsilon_{\mathcal{M}}=c_\mathcal{M}G\frac{m_{N_i}m_{N_j}}{P_N\cdot P_L}\int \frac{\rd^4 K}{(2\pi)^4} (K\cdot P_L) \imag \widetilde{\mathcal{M}}\,,
\end{equation}
where $G$ is the coupling defined below Eq.~\eqref{eq:zeroTepsilon} and $\widetilde{\mathcal{M}}$ is the so-called stripped amplitude introduced in Ref.~\cite{Seller:2024dkg}, that is the one-loop amplitude stripped of the couplings, the spinor structure, and the loop momentum ($K$) integration.
This formula holds for both the self-energy and the vertex correction diagrams shown in figure~\ref{fig:diagrams_combined} for both the $N_i\to \phi + L$ and the $\phi \to N_i + L$ decays with $\mathcal{M}$ being the corresponding amplitude at finite temperature. 
For a self-energy contribution $c_\mathcal{M}=4$, while for a vertex contribution $c_\mathcal{M}=2$.

\section{Thermal CP asymmetry factor}
\label{section:2}

The imaginary part of the amplitude required for the calculation of the thermal CP violating rate is evaluated using thermal field theory.
From the imaginary time formalism we know that each loop integration introduces one statistical factor, consequently at leading order the amplitudes are linear in $\fBF$ \cite{Bellac:2011kqa}.
However, the imaginary part of a diagram is readily given by the application of cutting rules that is more easily defined in the real time formalism of thermal field theory \cite{Kobes:1985kc,Kobes:1986za}.
All of these methods yield the same result \cite{Kobes:1990kr,Aurenche:1991hi}, although one has to define the amplitude carefully, as in the real time formalism the doubling of degrees of freedom results in an $n$-point function having $2^n$ components.
The so-called \emph{causal} $n$-point function, introduced by Kobes, is a combination of these components that is then linear in the statistical factor.
This causal combination is that which is obtained from the imaginary time formalism as well \cite{Aurenche:1991hi,Seller:2024dkg}.

The thermal CP asymmetry factor in the $N_i\to \phi + L$ and $\phi \to N_i + L$ decays receives contributions from four sources: from the single cut of the self-energy diagram ($\Sigma$) and from three cuts of the vertex correction diagram ($V$), as shown in figure~\ref{fig:diagrams_combined}.
The imaginary part of the decay diagram with the neutrino self-energy insertion evaluates to 
\begin{equation}
\label{eq:StrippedMselfenergy}
\begin{aligned}
    \imag&\widetilde{\mathcal{M}}^{\rm (\Sigma)}_{N_i\to \phi + L} = \imag\widetilde{\mathcal{M}}^{\rm (\Sigma)}_{\phi \to N_i + L} \\
    =&\, \frac{2\pi^2}{m_{N_i}^2-m_{N_j}^2}\big[1+\BE(p_N^0-k^0)-\FD(k^0)\big] \\
    &\times\bar\delta_L(K)\bar\delta_\phi(P_N-K)\,,
\end{aligned}
\end{equation}
where $\bar\delta_{\rm p}(P)\equiv \sgn(p^0)\delta(P^2-m_{\rm p}^2)$ for a generic momentum $P=(p^0,\p)$.
The difference in kinematics between the two decay processes results in $k^0>0$ for $N_i\to \phi+L$ whereas $k^0<0$ for $\phi\to N_i + L$.
Writing the statistical factors with positive arguments in the limit of $T\to 0$ leads to $\lim_{T\to 0}\imag\widetilde{\mathcal{M}}_{N_i\to\phi+L}^{(\Sigma)}$ being finite, whereas $\lim_{T\to 0} \imag\widetilde{\mathcal{M}}_{\phi\to N_i+L}^{(\Sigma)} = 0$.
The latter is trivially related to the vanishing decay width of $N_i$ when $m_\phi > m_{N_i}$ via the optical theorem.

The three cuts of the vertex diagram correspond to the three possible ways of selecting the two cut propagators from the three internal lines.
Denoting $K_2=K-P_N$ and $K_3=K-P_\phi$ with $K_1\equiv K$, we find
\begin{eqnarray}
\label{eq:strippedMvertex1}
%\begin{aligned}
&&\!\!\!\!\imag\widetilde{\mathcal{M}}_{N_i\to \phi + L}^{(V\text{, cut 1)}} = \imag \widetilde{\mathcal{M}}_{\phi\to N_i + L}^{(V \text{, cut 2)}} \\
&& =\pv\frac{2\pi^2}{K_3^2-m_{N_j}^2}\big[\FD(k_1^0)+\BE(k_2^0)\big]\bar \delta_L(K_1) \bar \delta_\phi(K_2)\,,
\nonumber
%\end{aligned}
%\end{eqnarray}
\\[2mm]
%\begin{eqnarray}
\label{eq:strippedMvertex2}
%\begin{aligned}
&&\!\!\!\!\imag\widetilde{\mathcal{M}}_{N_i\to \phi + L}^{(V\text{, cut 2)}} = \imag \widetilde{\mathcal{M}}_{\phi\to N_i + L}^{(V\text{, cut 1)}} \\
&&= \pv\frac{2\pi^2}{K_2^2-m_{\phi}^2}\big[\FD(k_1^ 0)-\FD(k_3^0)\big]\bar \delta_L(K_1) \bar \delta_{N_j}(K_3)\,,
\nonumber
%\end{aligned}
%\end{eqnarray}
\\[2mm]
%\begin{eqnarray}
\label{eq:strippedMvertex3}
%\begin{aligned}
&&\!\!\!\!
\imag\widetilde{\mathcal{M}}_{N_i\to \phi + L}^{(V\text{, cut 3)}} = -\imag \widetilde{\mathcal{M}}_{\phi\to N_i + L}^{(V\text{, cut 3)}} \\
&& = \pv\frac{2\pi^2}{K_1^2-m_{L}^2}\big[\BE(k_2^0)+\FD(k_3^0)\big]\bar\delta_\phi(K_2) \bar\delta_{N_j}(K_3)\,.
\nonumber
%\end{aligned}
\end{eqnarray}
Here $\pv$ denotes the principal value that is relevant for the loop momentum integration.
The cuts of the vertex correction are only formally equal for the two decay processes (up to a sign in case of the third cut), as momentum conservation implies different relations among the external momenta.

For the $N_i\to \phi + L$ decay the imaginary part of the amplitudes given in Eqs.~\eqref{eq:StrippedMselfenergy} and \eqref{eq:strippedMvertex1} can be extracted directly from the source term of lepton asymetry given in Eqs.~(57) and (71) of Ref.~\cite{Beneke:2010wd}, by writing them in the form given in Eq.~(44) of Ref.~\cite{Biondini:2017rpb} and using the form \eqref{eq:epsilon_M} for the CP asymmetry factor.
In the case of the vertex cuts 3 and 2 involving a Majorana neutrino, one can similarly use Eqs.~(14) and (19) of Ref.~\cite{Garbrecht:2010sz}.
For the $\phi \to N_i + L$ decay a similar crosscheck can be done with some extra effort based on Ref.~\cite{Frossard:2012pc}.

The imaginary parts of the stripped amplitudes in Eqs.~\eqref{eq:StrippedMselfenergy}--\eqref{eq:strippedMvertex3} are to be substituted into Eq.~\eqref{eq:epsilon_M}.
The integration over the loop momentum $K$ can be partially evaluated using the two Dirac delta distributions in each cut amplitude.
The remaining loop integration in the self-energy contribution can be evaluated in closed form, yielding an expression in terms of analytic functions (in particular logarithms and dilogarithms) \cite{Seller:2024dkg}.
The loop integrals for the vertex cuts can be calculated by choosing a suitable frame of reference.
In particular, for the second and third cuts it is most convenient to 
(i) choose the momentum of the external particle that is connected to the two on-shell propagators within the cut diagram as having $\p \parallel \hat{x}$, and 
(ii) use the energy of the same external particle as the free variable of the decay instead of the energy of the decaying particle \cite{Seller:2024dkg}.
This way each loop integral in the vertex cuts may be reduced to a single integral over the loop energy.

Once the loop integral in Eq.~\eqref{eq:epsilon_M} is evaluated, we are left to take the thermal average of the amplitude level CP asymmetry factor as indicated in Eq.~\eqref{eq:epsilonT_definition}.
Since for the second and third cuts of the vertex function the convenient external energy variable was not that of the decaying particle, we must rewrite the integral over the decaying energy in Eq.~\eqref{eq:epsilonT_definition} for an integral over the selected final state energy.
This change of variables is not trivial and it does not provide a one-to-one mapping of  the integral domain  \cite{Seller:2024dkg}.

The hierarchy and relative sizes of the masses of the particles involved can make for qualitatively different behavior of the thermal rate.
As customary in finite temperature calculations in cosmological settings, we take the mass squares of all particles to be given by a sum of their vacuum and thermal contributions \cite{Petitgirard:1991mf,Seller:2024dkg}.
In leptogenesis scenarios the vacuum masses of the SM particles are negligible as the relevant processes occur before the electroweak phase transition.
From the standard model we know that $m_\phi(T)>m_L(T)$ due to the large coupling between the Higgs field and the top quark.
It is conceivable that the Majorana mass is generated by a Higgs mechanism related to a new scalar field $\chi$ \cite{Chikashige:1980ui} and as a specific model we consider the Superweak extension of the Standard Model of Ref.~\cite{Trocsanyi:2018bkm}, in which case the thermal dependence of the vacuum expectation value of the scalar field is required \cite{Seller:2023xkn}. 
In this case we assume that the unstable Majorana neutrino is heavier than the SM particles at relatively low temperatures (vacuum mass term dominates when $\langle\chi\rangle\neq 0$), and is lighter at high temperatures (vacuum mass vanishes when $\langle\chi\rangle=0$).
In other phenomenological models the mass of the heavy Majorana fermion is taken to be constant, nevertheless this results in the same decay signature: $N_i\to \phi +L$ occurs at low temperatures while $\phi\to N_i+L$ occurs at high temperatures.

\begin{figure*}[!ht]
    \centering
    \includegraphics[width=0.75\linewidth]{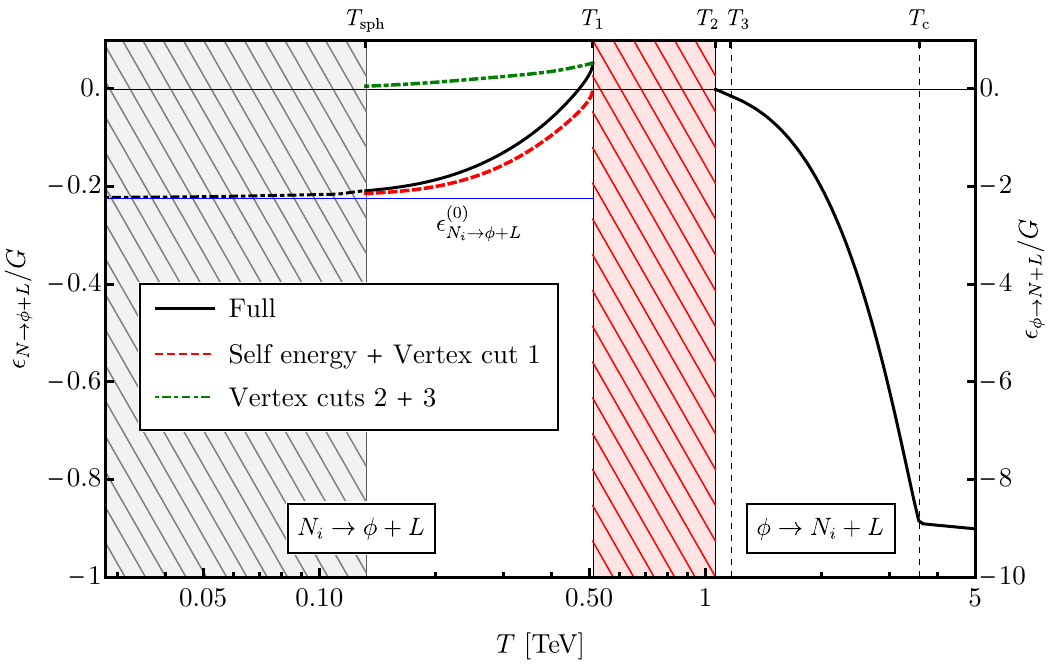}
    \caption{
    CP asymmetry factor in the {\it realistic mass scenario} of $N_i\to\phi+L$ and $\phi\to N_i+L$ decays.
    The solid black lines give our full result.
    At low temperature the contribution due to the purely thermal cuts of the vertex function is shown explicitly (green, dash-dotted).
    Above $T_{\rm c}$ the system is in the symmetric phase and $m_i\propto T$ for all particles.
    }
    \label{fig:main_plot}
\end{figure*}

The Majorana neutrino that appears in the loop cannot be identical to the external particle since for these diagrams $G=0$ (see Eq.~\eqref{eq:epsilon_M}).
In most phenomenological studies of leptogenesis one considers the CP asymmetry generated by the decays of the lightest Majorana neutrino.
Here we assume that there exist two flavors of Majorana neutrinos with the one in the loop being heavier than the external particle ($m_{N_i}<m_{N_j}$).
The relative difference between the neutrino masses is a crucial parameter of leptogenesis, in particular for resonant scenarios \cite{Pilaftsis:1997jf}.
We also mention that while at the leading order the vertex cuts involved in the $N_i\to \phi + L$ are always finite at $T>0$, the first and third cuts of the vertex function in the $\phi\to N_i + L$ decay vanish when $m_L + m_{N_i} \leq m_\phi \leq m_L + m_{N_j}$.

\section{Predictions for the CP asymmetry factor}

In this section we present the measure of CP violation in decays for two scenarios regarding the masses of the Majorana fermions.
In the \textit{realistic} case we consider a Lagrangian mass term $\mathcal{L}\supset y_{N_i}\chi N_i\Bar{N}_i^{\rm c}+\mathrm{h.c.}$ for the Majorana fermions with a singlet scalar field $\chi$ that develops a non-zero vacuum expectation value $w(T)$ below a critical temperature.
In the \textit{heavy} case we remain agnostic about the origin of the large Majorana mass and fix its value to be a constant.
In summary:
\begin{subequations}
\label{eq:mass_scenarios}
\begin{align}
\nonumber&\text{(i) }\textit{realistic mass scenario:} \\
&\quad m^2_{N_{i,j}}(T)=\frac{y_{N_{i,j}}^2}{2}w^2(T)+\frac{y_{N_{i,j}}^2}{16}T^2\,,
\\
\nonumber&\text{(ii) }\textit{heavy mass scenario:} \\
&\quad m_{N_{i,j}}(T)=m_{N_{i,j}}^{(0)}\,.
\end{align}
\end{subequations}
In scenario (i) we take the vacuum masses to be $m_{N_{i}}(T=0) \equiv m_{N_{i}}^{(0)} = 400\,$GeV and $m_{N_{j}}^{(0)} = 440\,$GeV, while in scenario (ii) the masses are fixed to $m_{N_i}^{(0)} = 1\,$TeV and $m_{N_j}^{(0)} = 100\,$TeV.
\red{Scenario (i) in particular is indicative of CP-asymmetry factors that may reproduce the observed baryon asymmetry through the mechanism of resonant leptogenesis \cite{Klaric:2021cpi}.}

As the CP asymmetry factor is a dimensionless quantity, it can only depend on dimensionless parameters.
The only scales present in the calculation are those of the neutrino masses and the temperature; therefore, the relevant dimensionless ratios are the mass ratio $m_{N_{i}^{(0)}}/m_{N_{j}^{(0)}}$ and the temperature-to-mass ratio $T/m_{N_{i,j}^{(0)}}$.
By fixing the mass ratio, the measure of CP violation remains approximately unchanged as a function of the dimensionless temperature $T/m_{N_{i}^{(0)}}$ (see Eq.~\eqref{eq:zeroTepsilon}, where the $T \to 0$ limit explicitly depends only on this ratio).

\begin{figure*}
    \centering
    \includegraphics[width=0.75\linewidth]{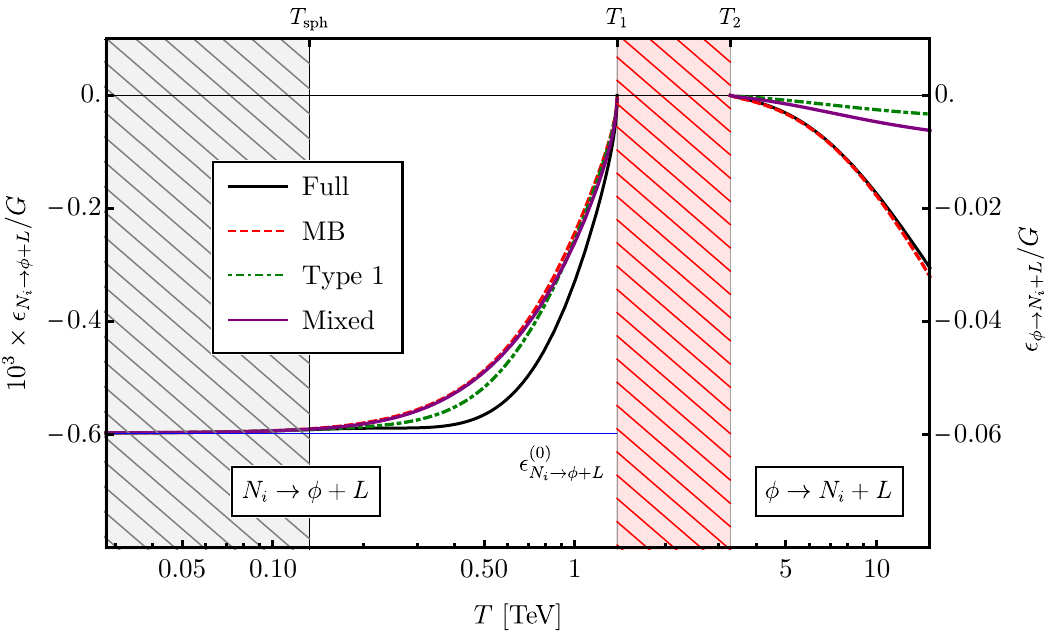}
    \caption{CP asymmetry factor in the {\it heavy mass scenario} of $N_i\to\phi+L$ and $\phi\to N_i+L$ decays.
    The solid black lines give our full result, while the others correspond to different predictions from the literature as explained in the main text.
    }
    \label{fig:constantMassPlot}
\end{figure*}

Figure~\ref{fig:main_plot} shows the temperature-dependent CP asymmetry factor in the realistic case, with 10\% relative mass gap between the heavy Majorana neutrinos.
The figure is separated into two parts by a kinematically excluded region (red hatched area) where neither $N_i\to\phi+L$ nor $\phi\to N_i+L$ are allowed. We emphasize that other decay and absorption processes as well as scatterings can take place in this region, and non-zero CP violation occurs here as well.
The indicated temperatures $T_i$ correspond to kinematic thresholds, $m_{N_i}(T_1)=m_{\phi}(T_1)+m_L(T_1)$, $m_{\phi}(T_2)=m_{N_i}(T_2)+m_L(T_2)$, and $m_{\phi}(T_3)=m_{N_j}(T_3)+m_L(T_3)$.
In the theory of baryogenesis via leptogenesis, the measure of CP violation is only relevant above the sphaleron freeze-out temperature, $T_{\rm sph}\simeq 132\,\GeV$ \cite{DOnofrio:2014rug}. 
There is no baryon number violation for $T\lesssim T_{\rm sph}$ (gray hatched area).

At $T\to 0$ our result approaches the vacuum value given by Eq.~\eqref{eq:zeroTepsilon}, assuming we ignore the vacuum masses of the SM fields, \ie, $m_\phi\sim m_L\propto T$.
For the $N_i\to \phi + L$ decay we have separated contributions due to the self-energy and the first cut of the vertex function from the second and third cuts of the vertex function.
The latter, purely thermal contributions are relevant at higher temperatures, close to the threshold temperature $T_1$ of the decay.
Coincidentally, these cuts should not be neglected when $m_{N_i}^{(0)}\simeq m_{N_j}^{(0)}$.
At high temperatures in the $\phi\to N_i+L$ the self-energy contribution dominates the result.
Nevertheless, note that the first and third cuts would only contribute at $T>T_3$.
At high temperature in the symmetric phase $w(T)=0$, so $m_{N_{i,j}}\propto T$. 
Consequently, the calculation becomes essentially scale-free and the measure of CP violation becomes roughly a constant (up to $T$-dependence through running couplings with renormalization scale $\mu=2\pi T$).

In figure~\ref{fig:constantMassPlot} we show the measure of CP violation for the heavy scenario where the masses of the Majorana neutrinos are kept fixed with a large relative mass gap.
\red{This figure is meant to indicate the qualitative behavior of the CP-asymmetry factor when calculated with heavy neutrinos and no thermal masses, although the particular choice made here likely does not correspond to a realistic scenario of leptogenesis.
As described in the paragraph starting below Eq.~\ref{eq:mass_scenarios}, this figure may be rescaled to show the CP-asymmetry factor for other values of $m_{N_{i,j}}$.}
The definitions of the regions and temperatures are the same as in figure~\ref{fig:main_plot}.
In both decays, the contributions with a cut internal neutrino propagator are exponentially suppressed, $\FD(T)\sim \exp(-m_{N_j}/T) \ll 1$.
The one-loop prediction to a good approximation is then the sum of the self-energy contribution and a single cut of the vertex function (topologically, cut 1 in the case of $N_i\to \phi + L$ and cut 2 in the case of $\phi \to N_i + L$).
In the limit $m_{N_j}\gg m_{N_i}$, the vacuum value given in Eq.~\eqref{eq:zeroTepsilon} reduces to a single term $\epsilon \simeq (3-1/x)\epsilon^{(\Sigma)}/2$.
This observation also holds for finite temperature, where the uncut $N_{j}$ propagator in the cut vertex diagram may be well-approximated by $-1/{m_{N_j}^2}$ (see also the comment below Eq.~(68) in Ref.~\cite{Beneke:2010wd}).

In figure \ref{fig:constantMassPlot}, we also compare our prediction with three others that may be found in the literature.
First, in the Maxwell-Boltzmann approximation (MB) we change all statistical factors to the MB distribution, $f_{\rm MB}(y) = \exp(-y)$.
Second, in the original Kobes-Semenoff approach to thermal field theory, the physically relevant thermal amplitudes were incorrectly identified as the component of the amputated $n$-point function that corresponded to all external legs being of ``type 1" (reflected by the label in the figure).
As a result, the one-loop diagrams contain also a term that is quadratic in the statistical factors, in contrast to the expectation in thermal field theory. 
This apparent inconsistency, pointed out in Refs.~\cite{Garny:2009qn,Garny:2009rv}, has been resolved in Ref.~\cite{Garny:2010nj} based on Refs.~\cite{Kobes:1990kr,Kobes:1990ua} by considering a specific, so-called ``causal'' combination of $n$-point functions containing ``type 1'' and ``type 2'' indices. 
This combination was also used in Ref.~\cite{Seller:2024dkg} to compute $\epsilon_{N_i\to\phi+L}$.
Third, in Ref.~\cite{Giudice:2003jh} the authors considered the thermal CP asymmetry factor by using the ``type 1" result and applying MB approximation to the final averaging over the external particle distributions.
Consequently, we dubbed this result as ``mixed" in the figure.

\section*{Summary}

In this letter we discussed the rate of lepton number violation in decay processes at finite temperature, which is a central quantity for leptogenesis scenarios based on the Boltzmann equation.
In particular, we presented the leading order perturbative estimate of the CP asymmetry factor in decays involving Majorana neutrinos. 
We demonstrated that the asymmetry is significantly enhanced when using a fully consistent calculation--one that includes all relevant cuts of the one-loop diagrams and employs the proper (causal) definition of the thermal amplitude.
We showed that in models where the masses of the heavy neutrinos are nearly degenerate, the purely thermal cuts of the vertex function become relevant. 
For hierarchical heavy neutrino masses, we also provided a comparison with existing estimates from the literature.
In particular, we found that using causal amplitudes becomes especially important at high temperatures.
Whether or not these effects will have significant influence on predictions of thermal leptogenesis in the Superweak extension of the Standard Model is subject to further work.

\section*{Acknowledgments}

This research was supported by the 
Excellence Programme of the Hungarian Ministry of Culture and Innovation under contract TKP2021-NKTA-64 and by the National Research, Development and Innovation Fund under contract ADVANCED 150794.
K.~S.~was partially supported by ÚNKP-23-4 New National Excellence Program of the Ministry for Culture and Innovation from the source of the National Research, Development and Innovation Fund, and also thanks the University of Hamburg and DESY for the kind hospitality during which parts of this letter were prepared.

%\bibliographystyle{elsarticle-num}

%\bibliography{refs}

\begin{thebibliography}{10}
\expandafter\ifx\csname url\endcsname\relax
  \def\url#1{\texttt{#1}}\fi
\expandafter\ifx\csname urlprefix\endcsname\relax\def\urlprefix{URL }\fi
\expandafter\ifx\csname href\endcsname\relax
  \def\href#1#2{#2} \def\path#1{#1}\fi

\bibitem{Fukugita:1986hr}
M.~Fukugita, T.~Yanagida, {Baryogenesis Without Grand Unification}, Phys. Lett.
  B 174 (1986) 45--47.
\newblock \href {https://doi.org/10.1016/0370-2693(86)91126-3}
  {\path{doi:10.1016/0370-2693(86)91126-3}}.

\bibitem{Biondini:2017rpb}
S.~Biondini, et~al., {Status of rates and rate equations for thermal
  leptogenesis}, Int. J. Mod. Phys. A 33~(05n06) (2018) 1842004.
\newblock \href {http://arxiv.org/abs/1711.02864} {\path{arXiv:1711.02864}},
  \href {https://doi.org/10.1142/S0217751X18420046}
  {\path{doi:10.1142/S0217751X18420046}}.

\bibitem{Laine:2022pgk}
M.~Laine,
{Sterile neutrino rates for general M, T, {\ensuremath{\mu}}, k: Review of a theoretical framework},
Annals Phys. \textbf{444}, 169022 (2022).
\newblock \href {http://arxiv.org/abs/2203.05772}{\path{arXiv:2203.05772}},
\href{https://doi.org/10.1016/j.aop.2022.169022}{\path{doi:10.1016/j.aop.2022.169022}}.

\bibitem{Klaric:2021cpi}
J.~Klari\'c, M.~Shaposhnikov and I.~Timiryasov,
{Reconciling resonant leptogenesis and baryogenesis via neutrino oscillations},
Phys. Rev. D \textbf{104} (2021) no.5, 055010.
\newblock \href {http://arxiv.org/abs/2103.16545}{\path{arXiv:2103.16545}},
\href {https://doi.org/10.1103/PhysRevD.104.055010}{\path{doi:10.1103/PhysRevD.104.055010}}.

\bibitem{Liu:1993tg}
J.~Liu, G.~Segre, {Reexamination of generation of baryon and lepton number
  asymmetries by heavy particle decay}, Phys. Rev. D 48 (1993) 4609--4612.
\newblock \href {http://arxiv.org/abs/hep-ph/9304241}
  {\path{arXiv:hep-ph/9304241}}, \href
  {https://doi.org/10.1103/PhysRevD.48.4609}
  {\path{doi:10.1103/PhysRevD.48.4609}}.

\bibitem{Xing:2011zza}
Z.-z. Xing, S.~Zhou, {Neutrinos in particle physics, astronomy and cosmology},
  Advanced Topics in Science and Technology in China, Springer, Berlin, 2011.

\bibitem{Covi:1997dr}
L.~Covi, N.~Rius, E.~Roulet, F.~Vissani, {Finite temperature effects on CP
  violating asymmetries}, Phys. Rev. D 57 (1998) 93--99.
\newblock \href {http://arxiv.org/abs/hep-ph/9704366}
  {\path{arXiv:hep-ph/9704366}}, \href {https://doi.org/10.1103/PhysRevD.57.93}
  {\path{doi:10.1103/PhysRevD.57.93}}.

\bibitem{Giudice:2003jh}
G.~F. Giudice, A.~Notari, M.~Raidal, A.~Riotto, A.~Strumia, {Towards a complete
  theory of thermal leptogenesis in the SM and MSSM}, Nucl. Phys. B 685 (2004)
  89--149.
\newblock \href {http://arxiv.org/abs/hep-ph/0310123}
  {\path{arXiv:hep-ph/0310123}}, \href
  {https://doi.org/10.1016/j.nuclphysb.2004.02.019}
  {\path{doi:10.1016/j.nuclphysb.2004.02.019}}.

\bibitem{Seller:2024dkg}
K.~Seller, Z.~Sz\'ep, Z.~Tr\'ocs\'anyi, {CP violation at finite temperature}
JHEP 09 (2025) 034
\newblock \href {http://arxiv.org/abs/2409.07180} {\path{arXiv:2409.07180}},  \href {https://doi.org/10.1007/JHEP09(2025)034}
  {\path{doi:10.1007/JHEP09(2025)034}}.

\bibitem{Beneke:2010wd}
M.~Beneke, B.~Garbrecht, M.~Herranen, P.~Schwaller, {Finite Number Density
  Corrections to Leptogenesis}, Nucl. Phys. B 838 (2010) 1--27.
\newblock \href {http://arxiv.org/abs/1002.1326} {\path{arXiv:1002.1326}},
  \href {https://doi.org/10.1016/j.nuclphysb.2010.05.003}
  {\path{doi:10.1016/j.nuclphysb.2010.05.003}}.

\bibitem{Anisimov:2010aq}
A.~Anisimov, W.~Buchm\"uller, M.~Drewes, S.~Mendizabal, {Leptogenesis from
  Quantum Interference in a Thermal Bath}, Phys. Rev. Lett. 104 (2010) 121102.
\newblock \href {http://arxiv.org/abs/1001.3856} {\path{arXiv:1001.3856}},
  \href {https://doi.org/10.1103/PhysRevLett.104.121102}
  {\path{doi:10.1103/PhysRevLett.104.121102}}.

\bibitem{Garbrecht:2018mrp}
B.~Garbrecht,
{Why is there more matter than antimatter? Calculational methods for leptogenesis and electroweak baryogenesis},
Prog. Part. Nucl. Phys. \textbf{110}, 103727 (2020).
\newblock \href {http://arxiv.org/abs/1812.02651}{\path{arXiv:1812.02651}},
\href {https://doi.org/10.1016/j.ppnp.2019.103727}{\path{doi:10.1016/j.ppnp.2019.103727}}.

\bibitem{Garny:2011hg}
M.~Garny, A.~Kartavtsev and A.~Hohenegger,
{Leptogenesis from first principles in the resonant regime},
Annals Phys. \textbf{328}, 26-63 (2013).
\newblock \href {http://arxiv.org/abs/arXiv:1112.6428}{\path{arXiv:1112.6428}},
\href{https://doi.org/10.1016/j.aop.2012.10.007}{\path{doi:10.1016/j.aop.2012.10.007}}.

\bibitem{Garny:2009qn}
M.~Garny, A.~Hohenegger, A.~Kartavtsev, M.~Lindner, {Systematic approach to
  leptogenesis in nonequilibrium QFT: Self-energy contribution to the
  CP-violating parameter}, Phys. Rev. D 81 (2010) 085027.
\newblock \href {http://arxiv.org/abs/0911.4122} {\path{arXiv:0911.4122}},
  \href {https://doi.org/10.1103/PhysRevD.81.085027}
  {\path{doi:10.1103/PhysRevD.81.085027}}.

\bibitem{Garny:2009rv}
M.~Garny, A.~Hohenegger, A.~Kartavtsev, M.~Lindner, {Systematic approach to
  leptogenesis in nonequilibrium QFT: Vertex contribution to the CP-violating
  parameter}, Phys. Rev. D 80 (2009) 125027.
\newblock \href {http://arxiv.org/abs/0909.1559} {\path{arXiv:0909.1559}},
  \href {https://doi.org/10.1103/PhysRevD.80.125027}
  {\path{doi:10.1103/PhysRevD.80.125027}}.

\bibitem{Garny:2010nj}
M.~Garny, A.~Hohenegger, A.~Kartavtsev, {Medium corrections to the CP-violating
  parameter in leptogenesis}, Phys. Rev. D 81 (2010) 085028.
\newblock \href {http://arxiv.org/abs/1002.0331} {\path{arXiv:1002.0331}},
  \href {https://doi.org/10.1103/PhysRevD.81.085028}
  {\path{doi:10.1103/PhysRevD.81.085028}}.

\bibitem{Kobes:1990kr}
R.~Kobes, {A Correspondence Between Imaginary Time and Real Time Finite
  Temperature Field Theory}, Phys. Rev. D 42 (1990) 562--572.
\newblock \href {https://doi.org/10.1103/PhysRevD.42.562}
  {\path{doi:10.1103/PhysRevD.42.562}}.

\bibitem{Garbrecht:2010sz}
B.~Garbrecht, {Leptogenesis: The Other Cuts}, Nucl. Phys. B 847 (2011)
  350--366.
\newblock \href {http://arxiv.org/abs/1011.3122} {\path{arXiv:1011.3122}},
  \href {https://doi.org/10.1016/j.nuclphysb.2011.01.033}
  {\path{doi:10.1016/j.nuclphysb.2011.01.033}}.

\bibitem{Frossard:2012pc}
T.~Frossard, M.~Garny, A.~Hohenegger, A.~Kartavtsev, D.~Mitrouskas, {Systematic
  approach to thermal leptogenesis}, Phys. Rev. D 87~(8) (2013) 085009.
\newblock \href {http://arxiv.org/abs/1211.2140} {\path{arXiv:1211.2140}},
  \href {https://doi.org/10.1103/PhysRevD.87.085009}
  {\path{doi:10.1103/PhysRevD.87.085009}}.

\bibitem{Aurenche:1991hi}
P.~Aurenche, T.~Becherrawy, {A Comparison of the real time and the imaginary
  time formalisms of finite temperature field theory for 2, 3, and 4 point
  Green's functions}, Nucl. Phys. B 379 (1992) 259--303.
\newblock \href {https://doi.org/10.1016/0550-3213(92)90597-5}
  {\path{doi:10.1016/0550-3213(92)90597-5}}.

\bibitem{Bellac:2011kqa}
M.~L. Bellac, {Thermal Field Theory}, Cambridge Monographs on Mathematical
  Physics, Cambridge University Press, Cambridge, 1996.

\bibitem{Kobes:1985kc}
R.~L. Kobes, G.~W. Semenoff, {Discontinuities of Green Functions in Field
  Theory at Finite Temperature and Density}, Nucl. Phys. B 260 (1985) 714--746.
\newblock \href {https://doi.org/10.1016/0550-3213(85)90056-2}
  {\path{doi:10.1016/0550-3213(85)90056-2}}.

\bibitem{Kobes:1986za}
R.~L. Kobes, G.~W. Semenoff, {Discontinuities of Green Functions in Field
  Theory at Finite Temperature and Density. 2}, Nucl. Phys. B 272 (1986)
  329--364.
\newblock \href {https://doi.org/10.1016/0550-3213(86)90006-4}
  {\path{doi:10.1016/0550-3213(86)90006-4}}.

\bibitem{Petitgirard:1991mf}
E.~Petitgirard, \emph{{Massive fermion dispersion relation at finite
  temperature}}, Z. Phys. C
  {\bfseries 54} (1992) 673.
  \newblock \href{https://doi.org/10.1007/BF01559497}
  {\path{doi.org/10.1007/BF01559497}}.
  
\bibitem{Chikashige:1980ui}
Y.~Chikashige, R.~N.~Mohapatra and R.~D.~Peccei, {Are There Real Goldstone Bosons Associated with Broken Lepton Number?},
Phys. Lett. B \textbf{98} (1981), 265-268.
\newblock \href {https://doi.org/10.1016/0370-2693(81)90011-3}{\path{doi:10.1016/0370-2693(81)90011-3}}.

\bibitem{Trocsanyi:2018bkm}
Z.~Tr\'ocs\'anyi, {Super-weak force and neutrino masses}, Symmetry 12~(1)
  (2020) 107.
\newblock \href {http://arxiv.org/abs/1812.11189} {\path{arXiv:1812.11189}},
  \href {https://doi.org/10.3390/sym12010107} {\path{doi:10.3390/sym12010107}}.

\bibitem{Seller:2023xkn}
K.~Seller, Z.~Sz\'ep, Z.~Tr\'ocs\'anyi, {Real effective potentials for phase
  transitions in models with extended scalar sectors}, JHEP 04 (2023) 096.
\newblock \href {http://arxiv.org/abs/2301.07961} {\path{arXiv:2301.07961}},
  \href {https://doi.org/10.1007/JHEP04(2023)096}
  {\path{doi:10.1007/JHEP04(2023)096}}.

\bibitem{Pilaftsis:1997jf}
A.~Pilaftsis, {CP violation and baryogenesis due to heavy Majorana neutrinos},
  Phys. Rev. D 56 (1997) 5431--5451.
\newblock \href {http://arxiv.org/abs/hep-ph/9707235}
  {\path{arXiv:hep-ph/9707235}}, \href
  {https://doi.org/10.1103/PhysRevD.56.5431}
  {\path{doi:10.1103/PhysRevD.56.5431}}.

\bibitem{DOnofrio:2014rug}
M.~D'Onofrio, K.~Rummukainen, A.~Tranberg, {Sphaleron Rate in the Minimal
  Standard Model}, Phys. Rev. Lett. 113~(14) (2014) 141602.
\newblock \href {http://arxiv.org/abs/1404.3565} {\path{arXiv:1404.3565}},
  \href {https://doi.org/10.1103/PhysRevLett.113.141602}
  {\path{doi:10.1103/PhysRevLett.113.141602}}.

\bibitem{Kobes:1990ua}
R.~Kobes, {Retarded functions, dispersion relations, and Cutkosky rules at zero
  and finite temperature}, Phys. Rev. D 43 (1991) 1269--1282.
\newblock \href {https://doi.org/10.1103/PhysRevD.43.1269}
  {\path{doi:10.1103/PhysRevD.43.1269}}.

\end{thebibliography}

\end{document}